\documentclass{jltp}

\usepackage{graphicx} 
\usepackage{amssymb}

\renewcommand{\vec}[1]{{\bf {#1}}}
\newcommand{\text}[1]{\mbox{#1}}
\newcommand{\w}{\omega}
\newcommand{\PP}{{\tilde{P}}}
\newcommand{\PPP}{{\tilde{P}_{N_0 M_0}}}

\newcommand{\sign}{\text{sign}}
\newcommand{\EWERT}[1]{\left\langle #1 \right\rangle}
\newcommand{\Ewert}[1]{\langle #1 \rangle}

\title{Optical Conductivity  and Pseudo-Momentum Conservation 
in Anisotropic Fermi Liquids\thanks{dedicated to Peter 
W\"olfle on the occasion of his 60th birthday}}

\author{Achim Rosch and Natan Andrei$^\dagger$}

\address{Institut f\"ur Theorie der Kondensierten Materie, Universit\"at Karlsruhe,\\
 D-76128 Karlsruhe, Germany\\
$^\dagger$Serin Laboratory,
Rutgers University, Piscataway, NJ 08855, USA}

\runninghead{A. Rosch and N. Andrei}{Pseudo-Momentum Conservation 
in  Anisotropic Fermi Liquids}

\begin{document}

\maketitle

\begin{abstract}
  Umklapp scattering determines the conductivity of clean metals.  In
  typical quasi one-dimensional Fermi liquids with an open Fermi
  surface, certain pseudo-momenta do not decay by  2-particle
  collisions even in situations where Umklapp scattering relaxes the
  momentum of the quasi particles efficiently.  Due to this
  approximate conservation of  pseudo-momentum, a certain fraction of
  the electrical current decays very slowly and a well-pronounced
  low-frequency peak emerges in the optical conductivity.  We develop
  simple criteria to determine under what conditions approximate
  pseudo-momentum conservation is relevant and calculate within  in
  Fermi liquid theory  the weights of the corresponding low-frequency
  peaks and the temperature dependence  of the various relevant decay
  rates. Based on these considerations, we obtain a qualitative
  picture of the frequency and temperature dependence of the optical
  conductivity of an anisotropic Fermi liquid.

\end{abstract}

\section{INTRODUCTION}
The transport properties of clean metals at finite temperatures are
determined by Umklapp scatterings which allow the electrical current
to decay by transfering momentum to the underlying crystal structure
in quantas given by the reciprocal lattice vectors $\vec{G}_i$
\cite{peierls,landau} .  The decay rates are  small for a clean metal
with a small Fermi surface and accordingly the conductivities can be
very large.  In the case of a large Fermi surface, momentum can relax
efficiently by Umklapp scattering. Quasi one-dimensional metals with
an open Fermi surface are a particularly interesting case, as certain
pseudo-momenta are still approximately conserved impeding the current
decay ``protecting''  a certain fraction of the current\cite{roschPRL}. 
This gives rise to well pronounced low-frequency
peaks in the optical conductivity as we will explain in detail below.

In this paper we shall introduce these approximately conserved ``pseudo-momenta'' and demonstrate that multi-particle processes are required to violate 
them. 
After a general discussion of the optical conductivity $\sigma(\w)$ in
the presence of the approximately conserved quantities, we will
investigate in detail under what conditions the decay of certain
pseudo-momenta is strongly suppressed. Finally, we study how these
pseudo-momenta influence the optical conductivity and the temperature
dependence of the resitivity.

\section{APPROXIMATE CONSERVATION LAWS AND OPTICAL CONDUCTIVITY}
\label{conduct.section}

We wish to study how weakly violated conservation laws and the resulting
 slowly decaying modes influence the optical
conductivity. Consider  a 
situation where a  quantity $\PP$
exists (to be refered to as  
``pseudo-momentum'')  with the following two properties:
i)  $\PP$ is approximately conserved  ii) the current $J$ 
has a finite cross-susceptibility
$\chi_{J \PP}$ with the pseudo-momentum. 

 We assume that the pseudo-momentum is
approximately conserved in the sense that its decay rate is slower than
any other quantity $Q$ with $\chi_{Q\PP}=0$ (``perpendicular'' to
$\PP$) and $\chi_{Q J}\neq 0$. In the following sections we will give
a number of examples of such pseudo-momenta, where the commutator
$[H,\PP] \approx 0$ vanishes for most of the relevant low-energy processes.
In particular we shall show that anisotropic Fermi Liquids possess such
approximate conservation laws.
The second requirement implies that
 a typical state with a finite expectation
value of the current will  also carry a finite pseudo-momentum and similarly
 a state with  finite pseudo-momentum will also be
characterized by a finite current: $J$ and $\PP$ have a finite
``overlap''. 


\begin{figure}
\centerline{\includegraphics[width=0.8 \linewidth]{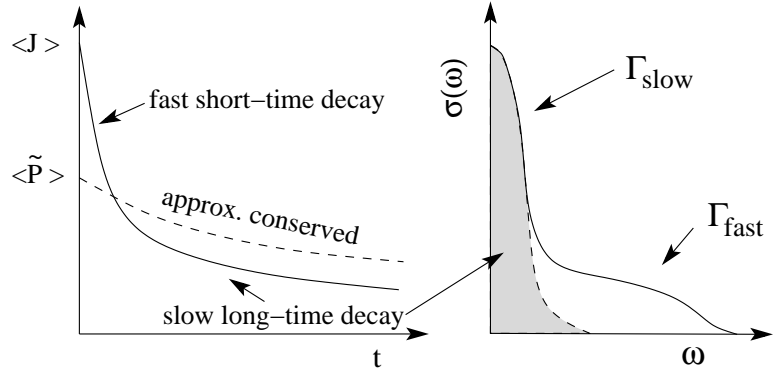}}
\caption[]{In a Gedanken experiment, a state with current  
  $\langle J \rangle$ is prepared. At time $t=0$, the
 external electric field is switched off. $J$ decays rapidly
  while the pseudo-momentum $\PP$ is approximately conserved and has a
  much lower decay rate $\Gamma_{\PP}$. In the optical conductivity
  $\sigma(\w)$ the approximate conservation of $\PP$ leads to a
  low-frequency peak with the width $\Gamma_{\PP}$ 
    and a height proportional to $1/\Gamma_{\PP}$. 
 In an anisotropic Fermi liquid, the width
   $\Gamma_{\PP}$ is given by Eqn. (\ref{gammaP}),
  the weight by (\ref{drudeRel}).
\label{figJdecay} }
\end{figure}

Now proceed with the following Gedanken experiment (see
Fig.~\ref{figJdecay}): prepare a state with a finite current
$\langle J(t=0) \rangle>0$ and switch off the driving electric field
at time $t=0$.  As the current is not conserved, it will decay rather
fast with a typical rate which we denote by $\Gamma_{J}$.  The
initial state with finite current will also have a finite pseudo
momentum $\langle \PP(t=0) \rangle$ with $\frac{\langle \PP(t=0)
  \rangle}{\langle J(t=0) \rangle }=\frac{\chi_{\PP J}}{\chi_{J
    J}}$  which will decay at a much slower rate than the current;
$\Gamma_{\PP}\ll \Gamma_{J}$. The slow decay of the pseudo-momentum will,
as time goes on, induce a slow  decay of the current since
 a state with finite pseudo-momentum will typically carry a
finite current $\langle J(t) \rangle= J(\langle \PP \rangle)\approx
\frac{\chi_{J \PP}}{ \chi_{\PP \PP}}\langle \PP(t) \rangle $.
Thus, in the long time limit, a finite fraction,
\begin{eqnarray}\label{drudeRel}
\frac{D}{D_0}=\frac{\chi_{J \PP}^2}{\chi_{\PP \PP}\chi_{J J} },
\end{eqnarray}
of $\langle J \rangle$ will not decay with the fast rate $\Gamma_{J}$
but with the much smaller rate $\Gamma_{\PP}$ as is shown
schematically in Fig.~\ref{figJdecay}.  

How does this affect the optical conductivity?  We argue
 (supported by rigorous  arguments\cite{mazur},
ana\-ly\-ti\-cal calculations\cite{roschPRL} and numerical
simulations\cite{garst}, see below) that a slow long-time decay of
some fraction of the current leads to a corresponding long-time tail
in the (equilibrium) current-current correlation function from which
the optical conductivity is calculated.  Therefore, one expects a
low-frequency peak in the optical conductivity which carries the
fraction $D/D_0$ of the total weight\cite{weight} $ \pi \chi_{J J}=2
\pi D_0=\pi \frac{n e^2}{m}$ with $\frac{n}{m}=\sum_{\vec{k \sigma}}
\frac{\partial^2 \epsilon_k}{\partial k^2} \langle
c^\dagger_{\vec{k}\sigma} c_{\vec{k}\sigma}\rangle $.

The width of the low-frequency peak is determined by the decay-rate of
$\PP$. As the weight of a peak is approximately given by its width
multiplied with its height, we expect that the height  and therefore
the dc-conductivity is of the order of $D/\Gamma_{\PP}$. The peak is
well pronounced if its height is well above the ``background'', i.e. if
\begin{eqnarray}\label{visible}
\frac{D}{\Gamma_{\PP}} \gg \frac{D_0}{\Gamma_{J}}.
\end{eqnarray}
In this case, the dc-conductivity is determined by the decay rate
$\Gamma_{\PP}$ of $\PP$.

A particular case is  the limit $\Gamma_{\PP}\to 0$, i.e.  when $\PP$
is {\em exactly conserved}. In this situation, studied long time ago
in a more general contex by Mazur\cite{mazur} and Suzuki\cite{suzuki},
the hand-waving arguments given above can be made rigorous. The
low-frequency peak evolves into a true  $\delta$-function and the
optical conductivity takes the form $\sigma(\w,T)= 2 \pi D(T)
\delta(\w) +\sigma_{\text{reg}}(\w,T)$ with a finite Drude weight
$D$. For a system with conserved charges $Q_n, n=1 \dots M,$  with $
\chi_{Q_n Q_m}=0$ for $n \ne m$,  the Mazur  inequality  reads,
\begin{eqnarray}\label{mazurInequ}
D\ge \frac{1}{2} \sum_{n=1}^M \frac{\chi_{J Q_n}^2}{\chi_{Q_n Q_n}}
\end{eqnarray}
Furthermore, Suzuki\cite{suzuki} showed that the inequality in
(\ref{mazurInequ}) can be re placed by an equality if the sum includes
{\em all} conservation laws! If therefore $\PP$ is the {\em only}
(approximately) conserved quantity 
in the system with a finite overlap
to the current $\chi_{J Q}\ne 0$ (as assumed above)
 then (\ref{drudeRel}) is exact as was tested numerically for
a simple model in Ref.~\onlinecite{garst}. The importance of the inequality for transport 
has been recently emphasized  by Zotos
{\it et al.}\cite{zotos}.

The above discussed low frequency peak in $\sigma(\w)$ with weight $D$
should not  be confused with the zero-temperature Drude weight
$D(T=0)$  as in general, $\lim_{T \to 0} D(T)$ need {\em not} be
identical to $ D(T=0)$. At low, but finite temperature, the low
frequency peak is well defined as $\Gamma_{\PP} \ll \Gamma_{J}
\frac{D}{D_0}$.  At $T=0$ in a clean metal, one expects
$\Gamma_{\PP}=\Gamma_{J}=0$ and therefore it is not possible to
extract the weight of the peak which is due to the slow mode
$\PP$. The same situation arises in finite temperature experiments when
the temperature is so low that $\Gamma_J$ is smaller than the energy
resolution.

\section{TRANSPORT IN AN ANISOTROPIC FERMI LIQUID}

\subsection{Pseudo-Momentum}
\label{pseudoFermiLiquid}

We introduce in this section pseudo-momentum operators  conserved 
by generic 2-particle low energy scattering terms. Since 
multi-particle or high-energy processes are
 required to violate their conservation laws the associated decay times
can be very long at low $T$.
 
We consider a three- or two-dimensional  anisotropic metal  with a
clearly defined most-conducting axis in $x$-direction.  It is assumed
that two well defined Fermi sheets perpendicular to this axis exist
(see Fig.~\ref{figFS}). The curvature of those sheets is not required
to be very small (see below). For simplicity, we discuss only
situations where a single band crosses the Fermi energy, but many of our
results can be generalized  to multi-band models. We consider a rather
arbitrary lattice, assuming only the existence of a translation vector
$\vec{a}_1$ of the underlying lattice in the x-direction, $\vec{G}_1$ is
the corresponding  reciprocal vector with $\vec{a}_1 \vec{G}_1=2 \pi$

\begin{figure}
\centerline{\includegraphics[width=0.9 \linewidth]{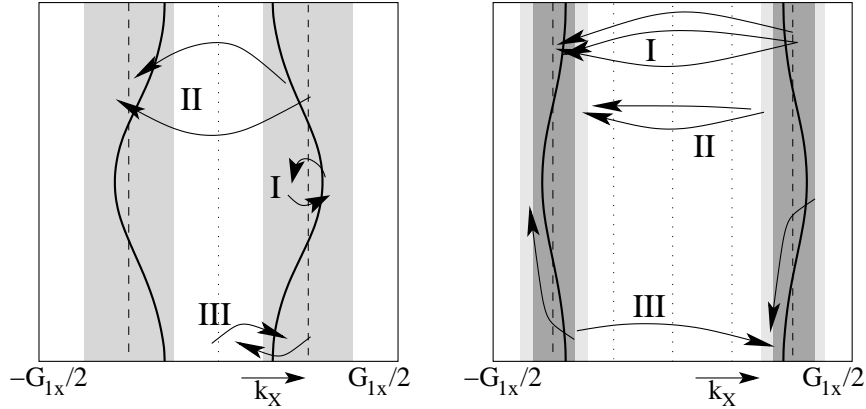}}
\caption{Left figure:
  Fermi surface (FS) of an anisotropic metal close to half filling.  Both
  ``forward'' (I) and ``Umklapp'' (II) scattering processes do not
  lead to an decay of the pseudo-momentum $\PP_{21}$ as long as the
  momenta are within the shaded area. The scattering event III
   leads to a decay of $\PP_{21}$ but is
  exponentially suppressed at low $T$ as it
  involves quasi particles far away from the Fermi surface.  Right
  figure: Filling close to
  $2/3$. Process I picks up the momentum $2 \vec{G}_1$ and 
relaxes both $P_x$ and $\PP_{21}$ but not
  $\PP_{32}$. The 2-particle scattering events of type II change $\PP_{32}$
  but are exponentially suppressed at low $T$ as long as the
  FS is in the shaded region. The 3-particle process III is only suppressed 
for a FS within the dark-shaded region (see Eqn.~(\ref{minN})).
 \label{figFS} }\end{figure}

Umklapp processes lead to a decay of any macroscopic momentum.  For
example, close to half filling, the process II shown in
Fig.~\ref{figFS} relaxes the momentum in $x$ direction via a momentum
transfer $G_1^x$. More generally, in the case of a filling fraction
close to  $M_0/N_0$ with integers $M_0$ and $N_0$,   the Fermi
momentum is approximately given by $k_F^x \approx \frac{M_0}{N_0}
\frac{G^1_x}{2}$ and therefore Umklapp processes where $N_0$ particles
move from one Fermi sheet to the other while picking up a lattice
momentum $M_0 G^1_x$ dominate the momentum relaxation for sufficiently
flat Fermi surfaces (see below).

We define now  a
pseudo-momentum  operator $\PPP$ with the property that it is conserved by
two-particle scattering processes close to the FS. It will decay only via 
high energy  two-particle  or multi-particle processes, and as a 
result  has 
a very slow decay. We term such quantities ``approximately conserved''. 
The pseudo-momentum is obtained by measuring the momentum in
$x$-direction on the left/right Fermi sheet with respect to the line
$k_x=\pm \frac{M_0}{N_0} G^1_x$ (dashed line in Fig.~\ref{figFS}),

\begin{eqnarray}
\tilde{P}_{N_0 M_0}=P_x-\frac{M_0}{N_0} \frac{G^1_x}{2} 
(N_R-N_L)=\sum_{\vec{k}} \delta k_{N_0 M_0} c^\dagger_{\vec{k}} c_{\vec{k}}
\end{eqnarray} 
where $N_{R/L}=\sum_{k_x \gtrless 0} c^\dagger_{\vec{k}} c_{\vec{k}} $ 
is the number of ``right-moving'' (left-moving) electrons
and, 
\begin{eqnarray}
\delta k_{N_0 M_0}=k_x-\frac{M_0}{N_0}\frac{G^1_x}{2} \text{sgn}(k_x).
\label{defineP}
\end{eqnarray}

To check to what extent $\PPP$ is conserved, consider 
 a generic 2-particle scattering term:
\begin{eqnarray}
H_2&=&\!\!\sum_{\text{1. BZ}} c^\dagger_{\vec{k}_1} c^\dagger_{\vec{k}_2}
 c_{\vec{q}_2} c_{\vec{q}_1}  V_{\vec{k}_1 \vec{k}_2,\vec{q}_1 \vec{q}_2} \sum_{\vec{G}_{\vec{n}}} \delta(\vec{k}_1+\vec{k}_2-
\vec{q}_1-\vec{q}_2-\vec{G}_{\vec{n}}) \nonumber 
\end{eqnarray}
and calculate the commutator of $\PP$ with it. We find,
\begin{eqnarray}
 [\PPP,H_2]&=&\!\!\sum_{\text{1. BZ}} c^\dagger_{\vec{k}_1} c^\dagger_{\vec{k}_2}
 c_{\vec{q}_2} c_{\vec{q}_1}  V_{\vec{k}_1 \vec{k}_2,\vec{q}_1 \vec{q}_2}
\sum_{\vec{G_{\vec{n}}}} \delta(\vec{k}_1+\vec{k}_2-\vec{q}_1-\vec{q}_2
-\vec{G}_{\vec{n}} ) \nonumber \\ 
&& \times (\delta k_{N_0 M_0}^{1}+\delta k_{N_0 M_0}^{2}-\delta {q^1_{N_0 M_0}}-
\delta {q^2_{N_0 M_0}})
\label{comm}
\end{eqnarray}
where the $\vec{G}_n$ are reciprocal lattice vectors.  For half
filling,  the right-hand side
of (\ref{comm}) vanish if all four momenta are in the shaded region of
Fig,~\ref{figFS}, i.e.  $|\delta k^{1}_{21}|$, $|\delta k^{2}_{21}|$,
$|\delta q^{1}_{21}|$, $|\delta q^{2}_{21}|< G_{1x}/4$:   For
``forward scattering'' processes of type I in Fig,~\ref{figFS},
$k_{1x}+ k_{2x}-k'_{1x}-k'_{2x}=0$,  both momentum $P_x$ and pseudo
momentum $\PP$ are conserved as for all 4 momenta have the same sign.
 ``Umklapp'' processes of type II pick up a lattice momentum
$G_x$ which is exactly compensated as two electrons are
moved from the right to the left Fermi surface due to the term
$\sign[k_x] G_{1x}/4$ in the definition of $\delta k_{21}$
(\ref{defineP}), and 
 $\PP$ is again conserved. The pseudo-momentum can
only decay by two-particle  high-energy  processes far from the Fermi surface,
e.g. III in Fig.~\ref{figFS}. Such
 scattering processes is exponentially suppressed even at moderate
temperatures. Note, that the usual version of the Boltzmann equation,
only 2-particle scattering processes are taken into account.
  
  Low-energy
  contributions can, however, result from these high-energy processes due to
  virtual excitations (in high orders of perturbation theory). Or to put 
it in the language of renormalization
  group: $N$-particle interactions are generated. Will they relax
  $\PPP$?  
Due to conservation of lattice momentum, 
$P_x$ can change only in quanta of $G^1_x$
(note that other reciprocal lattice vectors are perpendicular to $\vec{a}_1$)
and particle conservation guarantees that a change 
of $N_R-N_L$ will occur in multiples of $2$. 
Therefore $ \tilde{P}_{N_0 M_0}$ can be altered only by an amount 
$G^1_x \left(\Delta m + \Delta n \frac{M_0}{N_0}\right)$ with integers 
$\Delta m$ and $\Delta n$, i.e.
the smallest possible change of the pseudo-momentum is
$\Delta  \tilde{P}_{N_0 M_0}=\frac{1}{N_0} G^1_x$.
Therefore, a relaxation of $\PPP$ is not possible if all $2N$
  pseudo-momenta $\delta k_{ix}$ involved in an $N$-particle scattering process
  are smaller than $\frac{1}{2 N}\frac{1}{N_0} G^1_x$.  
 For a given Fermi surface, the decay of $\PP$
  by $N$-particle collision  at low $T$ is possible only  for
\begin{eqnarray}\label{minN}
N>N^*_{N_0 M_0}=\frac{G_{1x}/(2 N_0)}{ \max |\delta k^F_{N_0 M_0}|},
\end{eqnarray}
where $\max |\delta k^F_{N_0 M_0}|$  is the maximal distance of the
Fermi surface from the plane $k_x=\pm  \frac{M_0}{2 N_0} G_{1x}$
(dashed line in Fig.~\ref{figFS}). This simple geometrical criterion
together with its consequences discussed below, is the central result
of this paper. If one sets $N_0=1$ and
$M_0=0$ or $1$, one obtains a criterion for the decay of the momentum
$P_x$. Note that (\ref{minN}) is a necessary condition, there
can be situations, where it is not sufficient. 

At sufficiently high temperatures, the broadening of the Fermi-surface
and the thermal excitation of states with higher energy will favor
decay channels of the pseudo-momentum with smaller $N$ (this effect
can crudely be described by adding $T/v_F$ to 
$\max |\delta k^F_{N_0 M_0}|$
in (\ref{minN})).

We can at this point define precisely what is meant  by the
terms ``close to a commensurate filling''  or ``small curvature of the
Fermi surface'' in the context of this paper.  Given a Fermi surface and
a filling we wish to find  which
pseudo-momentum $\PP_{{ \tilde{N}_0 \tilde{M}_0}}$
will have the longest decay times at low $T$.
As processes with large $N$ are suppressed at low temperature due
to phase-space restrictions (see below), the answer can be obtained 
by determining the natural numbers 
$\tilde{N}_0$ and $\tilde{M}_0$ for which $N^*$ is maximal,
\begin{eqnarray}
\label{nStarTilde}
 N^*_{ \tilde{N}_0 \tilde{M}_0}=\max_{N_0, M_0} N^*_{N_0 M_0}.
\end{eqnarray}

In the right panel of  Fig.~\ref{figFS} the various relevant
scattering processes are discussed in a situation where  $\PP_{32}$ is
the most slowly decaying pseudo-momentum in the system. If the Fermi
surface is within the dark-shaded area (defined by setting $N=3$ on
the left-hand side of Eqn.~\ref{minN}), then  $\PP_{32}$ will decay neither
 by two- nor by three-particle scattering events.

In general, approximate conservation laws will be important 
 for any clean system at low temperatures if the relevant momentum or 
pseudo-momentum
{\em cannot} decay by the usual 2-particle processes, i.e.  if
\begin{eqnarray}
 N^*_{ \tilde{N}_0 \tilde{M}_0}>2.
\end{eqnarray}
For a generic  3-dimensional Fermi surface, $N^*_{ \tilde{N}_0
  \tilde{M}_0}$ will typically be smaller than 2: two-particle Umklapp
processes efficiently lead to a decay of the current and all relevant
momenta and pseudo-momenta. The situation is, however, different for
quasi one-dimensional metals with two well defined Fermi sheets with
moderate curvature and also for systems with a small FS. For example,
for $N_0=2$, it suffices if the FS is within the shaded region of
Fig.~\ref{figFS} as already discussed above.

If $\tilde{N}_0=1$, then the usual momentum $P_x$ is the most
important approximate conservation law (with $\tilde{M}_0=0$ for a
particle-like Fermi surface and  $\tilde{M}_0=1$ if a hole picture is
more appropriate). This will be the  case
for a  small density of particles (or holes) and  more generally
 for arbitrary filling if the Fermi
surface is closed. For an open Fermi surface, the factor $1/N_0$ in
(\ref{minN}) guarantees that for moderate curvature of the Fermi
surface only small values of $N_0$ are relevant. Large values of
$\tilde{N}_0$ can arise for incommensurate fillings 
in the extreme quasi one-dimensional 
limit, when $\delta k^F_{N_0 M_0}$ is very small.

\subsection{Decay Rates and dc-Conductivity}

The temperature dependence of the decay-rate $\Gamma_{\PPP}$ of $\PPP$
at low $T$ in the Fermi liquid regime is determined by the usual
phase-space arguments: a particle of energy $\w\sim T$ decays into
$2N^*-1$ particle and hole excitations, one of the energies is fixed by
energy conservation, and the remaining $2 N^*-2$ energies each have a
phase-space of order $\w$. Therefore,
\begin{eqnarray}
\Gamma_{\PPP} \propto T^{2 N-2}\label{gammaP}
\end{eqnarray}
where the integer $N$ is the smallest value consistent with
(\ref{minN}).  The prefactor in
(\ref{gammaP}) depends in a rather delicate way on the strength and
range of the interaction, the screening and the band-curvature. Note,
that a {\em local} $N$-particle interaction will give no contribution
for $N>2$ due to the Pauli-principle. Therefore the scattering rate is
strongly suppressed for weakly coupled chains with well-screened
interactions.  Furthermore, additional logarithmic temperature
dependences of the scattering vertices are expected even in the
Fermi-liquid regime as it is well known from Fermi liquid theory.

We want to stress that the analysis given above is valid for
interactions of arbitrary strength as long as a Fermi liquid
description is possible. For strong interaction, one should, however,
consider the pseudo-momentum of quasi particles which slightly
differs from the pseudo-momentum of the bare electrons (see below).

In section~\ref{conduct.section}, we have discussed how the decay rate
of the pseudo-momentum determines the dc-conductivity.  If
$\chi_{J\PP_{\tilde{N}_0 \tilde{M}_0}}^2/ \chi_{{\PP_{\tilde{N}_0
      \tilde{M}_0}\PP_{\tilde{N}_0 \tilde{M}_0}}}$ is finite for $T\to
0$  (which is the generic situation if the filling is not exactly commensurate
as we will discuss  in the next two sections),
then at lowest temperatures, the $T$-dependence of the dc-conductivity
of a clean Fermi liquid
will be determined by the decay rate of the pseudo-momentum and
therefore
\begin{eqnarray}\label{sigma}
\sigma(\w=0,T) \propto T^{-(2 N^*-2)}
\end{eqnarray}
where $N^*$ is the smallest integer larger than   $N^*_{ \tilde{N}_0
  \tilde{M}_0}$! This result is also valid in the case of a small
Fermi surface\cite{new}, where the well-known momentum $P_x$  is the
most important conservation law. In this case, it differs from the
often cited result by Peierls\cite{peierls,landau} $\sigma(T)\propto
T^{-2} e^{E_0/T}$. The latter result is obtained e.g. within the usual
Boltzmann equation, which takes into account only two-particle
scattering processes that are gapped by the energy $E_0$ for a small FS
and neglects higher order processes.  In many realistic situations,
even weak disorder or phonons will be the dominant relaxation
mechanisms of the pseudo-momenta and (\ref{sigma}) will not apply.  We
will argue below that, nevertheless, the approximate conservation of
pseudo-momentum still remains experimentally relevant.

As both the optical and the dc conductivity are determined not only by
decay rate of  pseudo-momentum but also by the weight of the
corresponding low-frequency peak, we will calculate the latter in the
next two sections. Then, one can use Eqn.~(\ref{visible}) to
determine, whether pseudo-momentum conservation is relevant and under
what conditions (\ref{sigma}) applies.

\subsection{Overlap of Current and Pseudo-Momentum: $\chi_{J_x \PP}$}

To obtain the weight of the low-frequency peak in $\sigma(\w,T)$, we
have to calculate according to (\ref{mazurInequ}) or (\ref{drudeRel})
the two susceptibilities $\chi_{J_x \PPP}$ and $\chi_{\PPP \PPP}$,
where $J_x$ is the electrical current in $x$ direction.
In this subsection we consider the overlap $\chi_{J_x \PPP}$.

 This overlap is almost completely fixed by current
conservation.  Static susceptibility can be calculated by taking
the limit $\w \to 0$ first and then $\vec{q}\to 0$ (for the
dc-conductivity the opposite limit is relevant). The continuity
equation for charge for $q_y,q_z=0, q_x \to 0$ reads
\begin{eqnarray}
\frac{\partial}{\partial t} \rho_{q_x}(t)+i q_x J_{x}(t,q_x)=0
\end{eqnarray} 
and the susceptibility is given by
\begin{eqnarray}
\chi_{J_x \PPP}&=&-i \lim_{q_x \to 0, \epsilon\to 0} \int_0^\infty dt 
e^{-\epsilon t} 
\EWERT{\left[J_{x}(t,q_x),\PPP(0,-q_x)\right]} \nonumber \\
&=& \lim_{q_x \to 0} \frac{1}{q_x} 
\EWERT{\left[\rho_{q_x}(t=0),\PPP(t=0,-q_x)\right]} \nonumber \\
&=& - e \sum_{\vec{k} \text{ in  1. BZ},\sigma}  \delta k_{N_0 M_0} 
\frac{\partial}{\partial k_x}
\EWERT{c^{\dagger}_{\vec{k}\sigma} c_{\vec{k}\sigma}} \nonumber \\
&=& e \Delta n_{N_0 M_0} + B_{\text{boundary}} \label{chiPJ}
\end{eqnarray}
where $\epsilon$ is infinitesimally small and  $\Delta n_{N_0
  M_0}=n-\frac{M_0}{N_0} n_{\text{max}}$, is the deviation of the
particle density from the filling $M_0/N_0$  (in units of particles
per volumem, $n_{\text{max}}$ corresponds to the density of 2 electrons per
unit cell). In the last step we used a partial integration which
produced boundary terms at the center and the edge of the first
Brillouin zone (BZ).  For the square BZ shown in Fig.~\ref{figFS} they
take the form
\begin{eqnarray}
B_{\text{boundary}}=  e G_{1x} \Bigl(
\sum_{{ \begin{array}{c}\vec{k} \\[-0.5ex]  k_x=0
  \end{array}}
  }  \frac{\EWERT{c^{\dagger}_{\vec{k}\sigma} 
c_{\vec{k}\sigma}}-1}{2} ~~~ + \!\!\!\sum_{{\begin{array}{c}\vec{k} \\[-0.5ex] 
k_x=\pm G_{1x}/2
  \end{array}}}
\!\!\!\!\!\! \frac{\EWERT{c^{\dagger}_{\vec{k}\sigma} 
c_{\vec{k}\sigma}}}{4}\Bigr). \nonumber
\end{eqnarray}
The boundary terms are obviously very small for an open Fermi surface:
deep in the Fermi sea, the occupation $\Ewert{c^{\dagger}_{k_x=0}
  c_{k_x=0}}$ is very close to 1, and at the edge of the Brillouin zone
it is almost 0. This is not true for strong interactions, but as
mentioned above,  we redefine $\PPP$ as the pseudo-momentum of the
quasi particles. The probability to find a quasi particle (hole) close
to the BZ boundary (at $k_x=0$) is exponentially small for the systems
discussed here.  Note that Luttinger's theorem guarantees that
$\Delta n_{N_0 M_0}$ for quasi particles is the same as for particles.
Therefore we conjecture that the boundary terms are exponentially
small,
\begin{eqnarray}\label{boundarySmall}
B_{\text{boundary}} \sim e^{-\beta \epsilon_F},
\end{eqnarray}
if $\PPP$ is appropriately defined.  It is easy to convince oneself
that this is true within Fermi liquid theory, we claim, however, that
(\ref{boundarySmall}) and (\ref{chiPJ}) hold under much more general
conditions.  The assumption, that boundary terms are exponentially
small and therefore negligible is for example implicitly assumed in
the standard derivation of the Luttinger liquid in one-dimensional
systems (in the continuum description which is the basis of the
Luttinger model, boundary contributions vanish exactly).

\subsection{$\chi_{\PPP \PPP}$ and Low-Frequency Weight in Fermi Liquid Theory}

The low-frequency weight $D$  also requires
$\chi_{\PPP \PPP}$ which can easily be calculated at low $T$ within Fermi
liquid theory following standard text books\cite{fermi}.  The result
will in general depend on the details of the momentum dependence of
the effective interactions and the band-structure. We assume a quasi
1D system with a Fermi velocity $v_F^*=k_F/m^*$ parallel to the most
conducting axis and, for simplicity, completely local interactions
characterized by two Fermi liquid parameter $F_{++}$ and $F_{+-}$ in
the spin-singlet channel to describe the interactions of two density
excitations $\delta n_{\vec{k}}$ on the same Fermi sheet or on
different sheets, respectively.  Then the relative
weight (\ref{drudeRel}) of  the low frequency peak in the optical
conductivity for low $T$ is given by
\begin{eqnarray} \label{drudeFL}
\frac{D}{D_0}\approx \frac{m}{m^*} \left(\frac{ 
\EWERT{(\delta k_{N_0 M_0})^2}_{\text{FS}}}
{ \EWERT{\delta k_{N_0 M_0}}^2_{\text{FS}}}-\frac{F_m}{1+F_m}\right)^{-1}.
\end{eqnarray}
$\EWERT{\dots}_{\text{FS}}$ is defined as an average over the Fermi
sheet, e.g. $\EWERT{\delta k_{N_0 M_0}}_{\text{FS}}=\int \!\!\! \int d
k_y d k_z (k_F^x-\frac{M_0}{N_0}\frac{G_x}{2})/(\int \!\!\! \int d k_y d k_z)$, where
$k_F^x=k_F^x(k_y,k_z)$ is the x-component of the Fermi momentum on the
right Fermi sheet, and  $F_m=F_{++}-F_{+-}$.  
Note that due to Luttinger's theorem $\Delta n_{N_0 M_0}=2
\EWERT{\delta k_{N_0 M_0}}_{\text{FS}}/(a_y a_z \pi)$, where $\Delta n_{N_0 M_0}$ is the
deviation of the  electron-density from half filling.

If the interactions are sufficiently weak so that no 
 phase transition is induced, the low-frequency weight $D$
vanishes close to half filling with
\begin{eqnarray}\label{drudeFLhalf}
\frac{D}{D_0} \sim \frac{m}{m^*} \left(\frac{\epsilon_F^*}{t_\perp^*}\right)^2
 \left(\frac{\Delta n_{N_0 M_0}}{n}\right)^2,
\end{eqnarray}
where $\epsilon_F^*=k_F v_F^*$ is the renormalized Fermi energy.  We
expect that the low-frequency weight $D$ decreases with increasing
temperature, mainly due to the thermal broadening of $\EWERT{(\delta
  k_x)^2}_{\text{FS}}$. Leading finite-$T$ corrections to
(\ref{drudeFL}) or (\ref{drudeFLhalf}) are of order
$(T/\epsilon_F^*)^2$.

\subsection{Qualitative Picture of the Optical Conductivity}
Combining the results of the previous sections, one can obtain a
qualitative picture of the low-temperature  optical conductivity in a
strongly correlated Fermi liquid. From the arguments given in this
paper, very little can be said about $\sigma(\w)$ at high frequencies
of the order of the (renormalized) Fermi energy, the behavior will in
general depend on details of the interactions and the band structure.
In a quasi one-dimensional situation and somewhat lower frequencies
(still above the scale where a Fermi liquid description is possible)
one might obtain results typical for a Luttinger  liquid with Umklapp
scattering\cite{giamarchi,roschPRL} (in the Luttinger liquid regime,
not discussed here, the conservation of pseudo-momenta can become
important\cite{roschPRL}).  In the Fermi liquid regime, which is at the
focus of this paper, two-particle collisions define the shortest
relevant time-scale and therefore one expects the well-know ``Drude''
peak in the optical conductivity with a width  $\Gamma\propto T^2$
given by the two-particle transport-rate, which we identify with the
fast decay rate $\Gamma_{J_x}$ of section~\ref{conduct.section} If
some pseudo-momentum  is approximately conserved and 
decays on a much longer time-scale at low
temperatures, we expect  a low-frequency peak in the optical
conductivity (sketched in Fig.~\ref{figJdecay}) on top of the much
broader $T^2$ ``Drude'' peak. Whether pseudo-momentum conservation is
important at lowest temperatures, depends on the structure of the
Fermi surface. To find the most relevant $\PPP$ in the limit $T\to 0$,
we determine $N^*_{N_0 M_0}$ defined in  Eqn.~(\ref{minN}) and
maximize it, to obtain the ``best conserved'' pseudo-momentum
$\PP_{\tilde{N}_0 \tilde{M}_0}$, see Eqn.~(\ref{nStarTilde}). If
$\chi_{ J \PP_{\tilde{N}_0 \tilde{M}_0}}\neq 0$, i.e. if
 the filling is {\em not} exactly
$\tilde{M}_0/\tilde{N}_0$ (see
Eqn.~(\ref{chiPJ})),  and if two-particle scattering processes
do not relax   $\PP_{\tilde{N}_0 \tilde{M}_0}$, i.e. $N^*_{\tilde{N}_0
  \tilde{M}_0}>2$, then a low-frequency peak is expected in the optical
conductivity of sufficiently clean samples.  Its  weight
 $D/D_0$ is given by Eqn.~(\ref{drudeFL}) and its
 width $\Gamma_{\PP_{\tilde{N}_0
    \tilde{M}_0}}$   by (\ref{gammaP}) and
therefore the dc-conductivity should be proportional to  $T^{-(2
  N^*-2)}$ at low $T$. The analysis given above was valid for very low (but finite) temperatures. For higher $T$, the 
``most relevant'' approximate conservation law can heuristically be 
determined by the following order of magnitude estimate for the dc-conductivity \begin{eqnarray}
\sigma(\w=0,T)\sim \max_{N_0, M_0} \!\left[
\frac{\chi_{J_x \PPP}^2}{\chi_{\PPP \PPP}} \frac{1}{\Gamma_{\PPP}}\right].
\end{eqnarray}
Methods for a more reliable calculation are shortly
discussed in the next section.

In many experimentally relevant situations, $\PP_{\tilde{N}_0
  \tilde{M}_0}$ will decay by scattering from impurities. If this
decay rate is sufficiently small --  see Eqn.~\ref{visible} --
pseudo-momentum conservation will still dominate $\sigma(\w,T)$ at low
frequencies. A typical situation in a weakly disordered  metal might
be the following: at very low temperatures, elastic scattering with a
rate $\Gamma_{el}$  due to disorder determines both current and
pseudo-momentum relaxation, therefore all pseudo-momenta are
irrelevant and $\sigma(\w=0)\sim D_0/\Gamma_{el}$.  The situation is
more interesting at somewhat higher temperatures, where inelastic
2-particle collisions dominate, $\Gamma_{el} < \Gamma_J \propto
T^2$. If $\Gamma_{el} \gg \Gamma_{\PP_{\tilde{N}_0 \tilde{M}_0}}$,
then the pseudo-momentum will decay by elastic scattering. In such a
situation, we expect a well defined low-frequency peak in the optical
conductivity (as shown in Fig.~\ref{figJdecay}) as long as the
inequality~(\ref{visible}) is fulfilled. In such a situation, the
dc-conductivity will be only weakly temperature dependent, and
$\sigma(\w=0)\sim D(T)/\Gamma_{el}$ as the 2-particle collisions are
not able to relax $\PP$.  Matthiessen`s rule is
obviously not valid.

\subsection{How to Calculate $\sigma(\w,T)$ Quantitatively}

As this paper  focuses on qualitative arguments, we have not tried to
calculate  the full frequency dependence of $\sigma(\w,T)$ for an
anisotropic Fermi liquid. Here, we discuss briefly the methods which
can be used to determine $\sigma(\w)$ quantitatively.

In the language of perturbation theory, the physics discussed in this
paper is the physics of vertex corrections. The presence of
approximate conserved quantities implies an approximate cancellation
of vertex- and self-energy  corrections at low frequencies. In any
perturbative calculation it is therefore essential to include the
proper vertex correction. This is done automatically, if one solves
the corresponding Boltzmann-type kinetic equation\cite{landau}. If,
however, the pseudo-momenta are conserved by 2-particle scattering
processes as discussed above, it is not sufficient to include only
2-particle processes in the collision term of the Boltzmann equation:
one has to consider collisions involving $N^*$ particles, where $N^*$
is the smallest integer larger than $N^*_{\tilde{N}_0 \tilde{M}_0}$
(\ref{nStarTilde}), to obtain the correct low-$T$ behavior. Similarly,
in perturbation theory, one has to include both self-energy and the
corresponding vertex corrections up to order $N^*$ in the interactions.

The fact that the  pseudo-momenta decay is much slower than other degrees
of freedom  in the system, suggests a hydrodynamic approach to
calculate $\sigma(\w)$.  This can be done with the help of the memory
matrix formalism  of Mori and Zwanzig\cite{morizwanzig,forster}.  In
the context of solid state physics, the use of this method has been
pioneered by G\"otze and W\"olfle\cite{woelfle}. The main input of
this method is the knowledge of the relevant slow variables, i.e. the
pseudo-momenta. Within the memory matrix approach, the relaxation
rates of those hydrodynamic variables is calculated
perturbatively. The main advantage is that one can avoid to solve
complicated transport- or vertex  equations. The correct weights are reproduced
if the relevant time scales are well seperated. For
Luttinger liquids, the memory matrix has been used  in Ref.~\onlinecite{roschPRL} to calculate $\sigma(\w)$ in situations where
pseudo-momenta are approximately conserved,  in
Ref.~\onlinecite{garst} this method has been compared with numerical
exact results for a certain classical model.

\section{CONCLUSIONS}

In this paper, we have discussed various approximate conservation laws
which determine the low-frequency conductivity of clean anisotropic
Fermi liquids with open Fermi surfaces. For a large class of
situations, the dominant scattering process is ineffective and does not
lead to a decay of
the current due to  the presence of some ``protecting'' 
pseudo-momentum which decays
on a much longer time-scale. In this situation,  the decay rate of the
pseudo-momentum determines the dc-conductivity. The most important
signature of this type of physics is a well-defined low frequency peak
in the optical conductivity as is shown schematically in
Fig.~\ref{figJdecay} with a weight which can be calculated from Fermi
liquid theory.

This paper has investigated the transport in  anisotropic Fermi
liquids.  The same physics  is relevant not only for Luttinger
liquids\cite{roschPRL}, but for example also quasi two-dimensional
d-wave superconductors with nodes close to
$\left(\frac{\pi}{2},\frac{\pi}{2}\right)$ as we will show in a future
publication.

\section{ACKNOWLEDGMENTS}
It is a honor and our great pleasure to dedicate this paper to
Prof. Peter W\"olfle on the occasion of his 60th birthday, especially
as he strongly influenced the present work in many discussions. We
would also like to thank M.~Garst, P. Prelov$\check{\text{s}}$ek and
X.~Zotos for valuable discussions.  This work was supported by the
Emmy-Noether program of the DFG.


\begin{thebibliography}{9}
\bibitem{weight} $D_0$ is the full optical weight for intraband
  processes, in a strongly correlated system, e.g. close to a Mott
  transition, it can be much larger than the weight in the two peaks
  in  Fig.~\ref{figJdecay} which are assumed to be due to scattering
  from well-defined quasi-particles.
\bibitem{new} In the case of a small Fermi surface where momentum
  conservation dominates, we do not expect this to be a new result but
  we were not able to find a related reference.
\end{thebibliography}
\end{document}